\documentclass[sigconf]{acmart}

\microtypesetup{expansion=false}
\setcopyright{none}
\acmYear{2027}
\copyrightyear{2027}
\acmISBN{}
\acmDOI{}

\title[Low-Rank Output Watermarks]{Watermark Evidence under Open-Weight LLM Modification: Exact Bounds for Low-Rank Output Watermarks}

\author{Prashant Shekhar}
\correspondingauthor
\affiliation{%
  \institution{Embry-Riddle Aeronautical University}
  \city{Daytona Beach}
  \state{FL}
  \country{USA}
}
\email{shekharp@erau.edu}

\author{Dumindu Samaraweera}
\affiliation{%
  \institution{Embry-Riddle Aeronautical University}
  \city{Daytona Beach}
  \state{FL}
  \country{USA}
}
\email{samarawg@erau.edu}

\author{Caroline Howard}
\affiliation{%
  \institution{Embry-Riddle Aeronautical University}
  \city{Daytona Beach}
  \state{FL}
  \country{USA}
}
\email{howac15@erau.edu}

\begin{abstract}
    Open-weight language models can continue generating after their released weights are modified, while a watermark verifier still scores outputs against the original base and marked models. We study how such modifications change the evidence available to an OpenStamp-style verifier. At each shared prefix, we represent the modification as a residual in the next-token logits. We derive an exact finite-vocabulary lower bound on the expected log-likelihood-ratio evidence under a residual-span constraint, along with a sharp span-only deviation bound. We also characterize the first-order evidence shift through a covariance term with an explicit second-order remainder, and determine the residual span needed to reach a target evidence level. For sequence-level detection, we derive an exact conditional exponential-moment bound and combine it with a supermartingale argument and a distribution-free order-statistic tolerance bound. This yields a high-confidence upper bound on miss probability after attack selection on calibration data and evaluation on independent certification continuations. In experiments with OpenStamp on Llama-2-7B and Mistral-7B-v0.3, INT8 quantization and 30\% pruning retain test true-positive rates of 85.6--94.8\%, while 500-step LoRA fine-tuning lowers them to 28.8\% and 11.6\%. The covariance prediction tracks the measured mean LoRA evidence drift on both models, with no remainder-bound violations across 160{,}000 test prefixes. Additionally, we observe zero false positives among 500 base-model continuations per model. Through these theoretical and empirical results, the current research provides a better understanding of watermark robustness to model-editing/tuning attacks.
\end{abstract}


\begin{document}
\maketitle

\section{Introduction}

Large language model (LLM) watermarking changes token probabilities so a detector can recognize a statistical signal in generated text \cite{kirchenbauer2023watermark}. Prior work studies statistical power, text editing, paraphrasing, and adaptive evasion \cite{zhao2024provable,pratihar2026catch22,diaa2025adaptive}. Open-weight release adds a model-level source of change, i.e., users can modify parameters or decoding, potentially changing the signal seen by a verifier. OpenStamp is a recent leading method in this setting; it embeds a structured low-rank output-head update and verifies text by comparing likelihoods under marked and privately retained base checkpoints \cite{bakshi2026openstamp}.
In OpenStamp, a fixed bank of green lists assigns watermark scores to vocabulary tokens. At each decoding step, the current decoder state determines weights over those lists, and the resulting vocabulary-wide bias is incorporated into a fixed update to the model's output head. Generation then uses the marked checkpoint as usual; verification compares the marked and base checkpoints' token probabilities and sums their log-likelihood ratios over the continuation. 

Our analysis uses this construction as its motivating example, but the evidence bounds do not require the specific \(GSP\) factorization. They also apply to other weight-embedded watermarks when verification uses paired base/marked next-token laws and each post-release change admits a predictable bound on its logit-residual span, as formalized in Section~\ref{sec:model}.
Let \(\mathcal V\) be the generated-token vocabulary, \(V=|\mathcal V|\) its size, \(d_{\mathrm{hid}}\) the decoder-state width, and \(L\) the number of fixed green lists. The matrix \(G\in\mathbb R^{V\times L}\) contains one green-list mask per column, with each green entry scaled by watermark strength \(\delta_{\mathrm{wm}}>0\); \(\gamma\in(0,1)\) is the green-list fraction. Let \(d_{\mathrm{proj}}\) be the projection width, with \(P\in\mathbb R^{d_{\mathrm{proj}}\times d_{\mathrm{hid}}}\) and \(S\in\mathbb R^{L\times d_{\mathrm{proj}}}\), so \(SP\in\mathbb R^{L\times d_{\mathrm{hid}}}\). At decoding position \(t\), \(h_t\in\mathbb R^{d_{\mathrm{hid}}}\) is the decoder state and \(s_t=SP h_t\in\mathbb R^L\) is its vector of list weights. Let \(\Delta W\) denote the fixed output-head update and \(d_t\) its induced vocabulary-wide logit shift at position \(t\). OpenStamp's factorization is
\begin{equation}
  \Delta W=GSP,\qquad d_t=\Delta W h_t=G s_t,
  \label{eq:gsp}
\end{equation}
where \(\Delta W\in\mathbb R^{V\times d_{\mathrm{hid}}}\) is the fixed output-head update, \(P\) projects the decoder state, and \(S\) selects list directions. Thus \(\operatorname{rank}(\Delta W)\le L\). The watermark is carried by the released weights; verification uses the original base/marked pair.

At a shared prefix, a post-release modification changes the logits by a residual vector. The residual's span controls the largest possible redistribution of token probability, while its alignment with the watermark scores determines the first-order change in expected evidence. We quantify both effects, carry the prefix-level bounds through to sequence-level miss risk, and test their predictions under modifications selected on calibration data and evaluated on independent continuations. Broadly, the contributions of the current research can be summarized as follows:

\begin{itemize}
  \item We derive an exact finite-breakpoint lower envelope for the conditional reference-LLR mean under a bounded logit-span residual, together with a sharp deviation bound determined by the ranges of the watermark logit shift and the modification residual.
  \item We invert the envelope to obtain the exact per-prefix span required to reach a target mean score. The minimizing two-level residual has a constructive minimum-Frobenius rank-one output-head realization at that prefix.
  \item We derive an exact conditional exponential-moment envelope and combine it with a distribution-free order-statistic tolerance limit to bound miss probability after calibration-based attack selection.
  \item We evaluate Llama-2-7B and Mistral-7B-v0.3 under INT8 quantization, 30\% magnitude pruning, and 500-step OpenWebText LoRA fine-tuning. INT8 and pruning retain true-positive rates (TPRs) of 85.6--94.8\%; LoRA lowers them to 28.8\% and 11.6\%. On 160,000 test prefixes, the covariance expansion tracks the direction and scale of mean evidence drift; all eight familywise 95\% holdout bounds remain 1.0.
\end{itemize}

We first define the paired-model score and attacker, derive prefix-level evidence bounds and a sequence-level risk certificate, and evaluate them under three post-release modifications. Section~\ref{sec:related-work} reviews prior work on text watermarks, open-weight model modification, and statistical certification.

\section{Related Work}\label{sec:related-work}

\paragraph{Text watermarking, detection, and evasion.}
Green-list schemes encode a detectable signal by changing next-token probabilities \cite{kirchenbauer2023watermark}. Later work studies error rates, distribution-adaptive design, and robustness to text edits \cite{hu2024unbiased,he2025dawa,cai2026statistical,zhao2024provable,li2026trgof,pratihar2026catch22}; adaptive paraphrasing attacks target known watermark procedures \cite{diaa2025adaptive,huang2026rlcracker}.

\paragraph{Open-weight and model-embedded watermarks.}
Open-weight methods protect model ownership or mark generated text. Quantization-dependent parameter marks target ownership \cite{li2023quantization}; WAPITI and GaussMark embed signals in weights \cite{chen2024wapiti,block2025gaussmark}. MarkTune and PRO train text marks for quality or downstream robustness \cite{zhao2025marktune,xue2026pro}; Gloaguen et al. test durability under transformations and Scharff et al. address merging \cite{gloaguen2025durability,scharff2026merge}. Christ et al. give parameter-robustness guarantees under explicit attacker assumptions \cite{christ2024provably}. OpenStamp's fixed low-rank output update and paired-model LLR motivate our setting \cite{bakshi2026openstamp}; Mark Your LLM instead targets model-IP violations and text misuse \cite{xu2025markyourllm}. We fix the OpenStamp verifier and bound evidence change under checkpoint modification.

\paragraph{Watermark certification and statistical bounds.}
Randomized and mollifier smoothing certify parameter-embedded neural-network marks against parameter perturbations \cite{bansal2022certified,ren2023dimension}; Feng et al. apply smoothing to LLM text-watermark detection under text changes \cite{feng2024certified}. Unbiased Watermark optimizes a likelihood-ratio score over a total-variation neighborhood \cite{hu2024unbiased}. We bound logit-residual span and combine density-ratio and Hilbert-metric arguments, supermartingale Chernoff bounds, and a maximum-order-statistic tolerance limit \cite{benavoli2013densityratio,cohen2023hilbert,howard2020chernoff,wilks1941tolerance}.

\section{Model}\label{sec:model}

\subsection{Paired models and detector}

Fix a prompt context \(c\) and a scored continuation length \(N\). Let \(Y_{1:N}\) be the random continuation, and write \(Y_{<t}=(Y_1,\ldots,Y_{t-1})\) for the random prefix and \(y_{<t}=(y_1,\ldots,y_{t-1})\) for a realized prefix. At position \(t\), let \(z_t\in\mathbb R^V\) be the base-model logits and \(d_t\in\mathbb R^V\) the fixed watermark logit shift defined in Eq.~\ref{eq:gsp}. For \(x\in\mathbb R^V\), define \(\operatorname{softmax}(x)(v)=e^{x(v)}/\sum_{u\in\mathcal V}e^{x(u)}\). Then
\[
 p_t=\operatorname{softmax}(z_t),\qquad
 q_t=\operatorname{softmax}(z_t+d_t).
\]
We assume generation and scoring use these same categorical probabilities, with common support. The reference token score and sequence score are
\begin{equation}
 \ell_t(v)=\log\frac{q_t(v)}{p_t(v)}
 =d_t(v)-\log\!\left(\sum_u p_t(u)e^{d_t(u)}\right).
 \label{eq:llr}
\end{equation}
For categorical distributions \(q\) and \(p\) on \(\mathcal V\), we use \(D_{\mathrm{KL}}(q\Vert p)=\sum_{v\in\mathcal V}q(v)\log(q(v)/p(v))\) for Kullback--Leibler divergence.
At position \(t\), define the mean reference evidence under the original marked law by \(k_{1,t}=D_{\mathrm{KL}}(q_t\Vert p_t)=\mathbb E_{q_t}[\ell_t]\).
The fixed-horizon detector score is
\[
 \Lambda_N=\sum_{t=1}^N\ell_t(Y_t).
\]
At threshold \(\tau\), the detector declares marked text when \(\Lambda_N\ge\tau\). Let \(\Pr_0\) denote probability when tokens are sampled from the base conditionals \(p_t\), and \(\Pr_{\widetilde q}\) probability under the modified-marked conditionals \(\widetilde q_t\). The false-positive and miss probabilities are \(\alpha_N(c,\tau)=\Pr_0(\Lambda_N\ge\tau)\) and \(\widetilde\beta_N(c,\tau)=\Pr_{\widetilde q}(\Lambda_N<\tau)\). We call these the false-positive rate (FPR) and false-negative rate (FNR), respectively; the true-positive rate (TPR) is \(1-\widetilde\beta_N(c,\tau)\). Since \(e^{\Lambda_N}\) is the marked-to-base sequence likelihood ratio, \(\mathbb E_0e^{\Lambda_N}=1\), so Markov's inequality gives \(\alpha_N(c,\tau)\le\min\{1,e^{-\tau}\}\).

At a shared token prefix, represent a post-release change by its conditional logit residual \(r_t\). For any vector \(x\in\mathbb R^V\), define its vocabulary span by \(\operatorname{osc}(x)=\max_{v\in\mathcal V}x(v)-\min_{v\in\mathcal V}x(v)\). Then
\begin{equation}
 \widetilde q_t=\operatorname{softmax}(z_t+d_t+r_t).
 \label{eq:modified}
\end{equation}
For vectors \(x,y\), \(xy^\top\) denotes their outer product; for a vector \(x\in\mathbb R^V\), \(\operatorname{diag}(x)\) is the diagonal matrix with entries \(x(v)\). We use \(\langle x,y\rangle=x^\top y\), \(\|x\|_2=(\sum_jx_j^2)^{1/2}\) for vectors, and \(\|A\|_F=(\sum_{i,j}A_{ij}^2)^{1/2}\) for matrices. For operations that change internal layers, \(r_t\) is the transformed-minus-reference logit vector on the same prefix. An output-head update \(A\in\mathbb R^{V\times d_{\mathrm{hid}}}\) gives \(r_t=Ah_t\). Its row diameter is \(D_A=\max_{u,v\in\mathcal V}\|A_{u,:}-A_{v,:}\|_2\), where \(A_{u,:}\) denotes the row for token \(u\). If \(\|h_t\|_2\le H\), where \(H\) is a bound on the state norm, then \(\operatorname{osc}(r_t)\le H D_A\). For the general theory, \(R_t(c,y_{<t})\ge0\) is a predictable budget, determined by the prompt and prefix before \(Y_t\) is drawn, that satisfies \(\operatorname{osc}(r_t)\le R_t(c,y_{<t})\). A common shift of all logits has span zero and leaves the softmax unchanged.

\subsection{Threat model and assumptions}

The attacker receives the marked checkpoint and may choose a transformation using calibration data. The verifier retains the original base and marked models and scores text with \(\Lambda_N\). For the finite-sample result, the chosen transformation, threshold, Chernoff parameter \(\lambda<0\), and predictable budgets are frozen before an independent certification sample is drawn. At every history, the budget \(R_t(c,y_{<t})\) must bound \(\operatorname{osc}(r_t)\). Applying the guarantee to a transformation therefore requires a valid bound over the histories covered by certification.

In the empirical study, the evaluator has each transformed checkpoint and computes its exact same-prefix logit residual. Deployment without access to that checkpoint requires a certified upper bound on residual span; the measured spans here serve as evaluation diagnostics.

The theory uses a fixed context and horizon. For a prompt population, the same argument applies with independently sampled contexts included in the sequence-level law. Experiments use raw temperature-1 softmax sampling for exactly 64 steps. Other decoding policies require the generation law and verifier score to include the same policy.

\section{Results}

\subsection{From logit change to one-step evidence}

Fix a position and suppress its subscript \(t\). The verifier assigns score \(\ell(v)\) to token \(v\), while the modified marked checkpoint samples from \(\widetilde q\). The first question is how low the expected reference score can fall under a residual-span budget \(R\). Write \(p=\operatorname{softmax}(z)\), \(q=\operatorname{softmax}(z+d)\), \(\widetilde q=\operatorname{softmax}(z+d+r)\), \(k_1=D_{\mathrm{KL}}(q\Vert p)=\mathbb E_q[\ell]\), and \(\widetilde\mu=\mathbb E_{\widetilde q}[\ell]\).

\begin{theorem}[Exact mean envelope and sharp span bound]\label{thm:envelope}
For $u\in[0,1]$, define the interpolating distribution
\[
 q_u=\operatorname{softmax}(z+d+ur)
 \quad\text{and}\quad
 \Sigma_u=\operatorname{diag}(q_u)-q_uq_u^\top.
\]
Thus, \(\Sigma_u\) is the covariance matrix of a categorical draw from \(q_u\).
For every residual $r$,
\begin{equation}
 \widetilde\mu-k_1=\langle\widetilde q-q,d\rangle
 =\int_0^1d^\top \Sigma_u r\,du,
 \label{eq:drift}
\end{equation}
For a span budget $R\ge0$, set $\rho=e^R$, order tokens $v_1,\ldots,v_V$ so that $\ell(v_1)\le\cdots\le\ell(v_V)$, and define
\[
 Q_j=\sum_{i\le j}q(v_i),\qquad
 C_j=\sum_{i\le j}q(v_i)\ell(v_i),\qquad Q_0=C_0=0.
\]
Then the exact worst conditional mean is
\begin{equation}
 \underline\mu(R)=
 \min_{0\le j\le V}
 \frac{k_1+(\rho-1)C_j}{1+(\rho-1)Q_j}.
 \label{eq:mean-envelope}
\end{equation}
The minimum is attained by a residual that takes one value on an initial segment of the ordered tokens and another value on its complement. Also, for every residual,
\begin{equation}
 |\widetilde\mu-k_1|
 \le \operatorname{osc}(d)\tanh\!\left(\frac{\operatorname{osc}(r)}4\right)
 \le \frac{\operatorname{osc}(d)\operatorname{osc}(r)}4.
 \label{eq:tanh}
\end{equation}
The first inequality is sharp given only the two spans.
\end{theorem}

The integral in~\eqref{eq:drift} tracks how token probabilities move as the modification is applied.

Let \(X\) be a token-valued random variable with law \(q\). For functions \(a,b:\mathcal V\to\mathbb R\), write \(\operatorname{Cov}_q(a(X),b(X))=\mathbb E_q[a(X)b(X)]-\mathbb E_q[a(X)]\mathbb E_q[b(X)]\) for their covariance under \(q\).
Intuitively, evidence drops when a modification shifts probability toward tokens carrying less watermark evidence. The next corollary quantifies that alignment and bounds the error in its first-order estimate.
\begin{corollary}[Directional first-order response]\label{cor:directional}
For every residual $r$, there is a remainder $\varepsilon$ such that
\begin{equation}
 \widetilde\mu-k_1=\operatorname{Cov}_{q}(d(X),r(X))+\varepsilon,
 \qquad
 |\varepsilon|\le\frac18\operatorname{osc}(d)\operatorname{osc}(r)^2.
 \label{eq:directional-expansion}
\end{equation}
Thus the covariance gives the first-order change in expected reference evidence, and the displayed bound controls the error for a finite residual.
\end{corollary}

The covariance is positive when the residual tends to raise logits on high-scoring tokens and lower logits on low-scoring tokens under $q$; it is negative when the residual shifts probability in the opposite direction. A small covariance leaves a second-order change controlled by~\eqref{eq:directional-expansion}. The exact envelope still optimizes over every residual with the allowed span and uses the full token probabilities and scores; its minimizer raises relative mass on the lowest-scoring tokens. Sorting costs $O(V\log V)$ time and $O(V)$ memory. The $\tanh$ bound uses only the two spans and can be looser. All these bounds apply prefix by prefix; a shared parameter update can induce different residuals along different histories.

The envelope maps a span budget to its worst expected score. The next result inverts that map: it gives the least span that reaches a target $\eta$ and constructs an output-head update that realizes the extremum at one prefix.

\begin{corollary}[How Much Logit Change Is Needed to Reach a Target Evidence Level?]\label{cor:breakdown}
For a target $\eta$ with $\min_v\ell(v)<\eta<k_1$, the smallest span at which the worst mean score reaches $\eta$ is
\begin{equation}
 R^\star(\eta)=
 \log\!\left(1+
 \min_{j\ge1:\ C_j/Q_j<\eta}
 \frac{k_1-\eta}{\eta Q_j-C_j}\right).
 \label{eq:radius}
\end{equation}
Let \(j^\star\) be a minimizing index in~\eqref{eq:radius}. The corresponding two-level residual attains mean score $\eta$. At a prefix with nonzero hidden state $h$, center this residual to $\bar r^\star$ by subtracting its vocabulary mean and set
\[
 A^\star=\frac{\bar r^\star h^\top}{\|h\|_2^2}.
\]
Then $A^\star h=\bar r^\star$, $\operatorname{rank}(A^\star)\le1$, and the output-head update $A^\star$ induces the same conditional distribution as the least-favorable residual. Its minimum Frobenius norm, with the lower score level assigned to the \(j^\star\) lowest-scoring tokens, is
\[
 \|A^\star\|_F=
 \frac{R^\star(\eta)}{\|h\|_2}\sqrt{\frac{j^\star(V-j^\star)}{V}}.
\]
This construction realizes the least-favorable law at one prefix. Across a sequence, the residual induced by a shared matrix varies with the history.
\end{corollary}

The radius translates a target score into a local span threshold. The minimum-norm construction realizes the extremum at that prefix; its norm scales inversely with the hidden-state norm.

\subsection{From one-step evidence to a sequence-level risk bound}

The mean envelope describes expected evidence at one prefix. To bound sequence-level miss probability for the fixed threshold, we also need the conditional moment of $e^{\lambda\ell}$ for $\lambda<0$. The next theorem optimizes a general token payoff over the same bounded-span family and gives this moment exactly.

At a fixed prefix, let $q$ denote the original marked next-token distribution. For a payoff $f:\mathcal V\to\mathbb R$, let $m_f=\mathbb E_q f$. For $\theta\in[0,1]$, define the largest payoff mass that can be assigned to a $q$-mass $\theta$ by
\[
 U_f(\theta)=
 \max_{\substack{0\le u_v\le1\\\sum_v q(v)u_v=\theta}}
 \sum_vq(v)u_v f(v).
\]
Here $U_f(\theta)$ is the largest total payoff carried by a $q$-mass of size $\theta$. Its maximizing allocation fills tokens in decreasing order of $f$.

\begin{theorem}[Worst-Case Evidence Moment under a Bounded Logit Change]\label{thm:mgf}
For every span budget $R$, with $\rho=e^R$, the maximum of $\mathbb E_{\operatorname{softmax}(z+d+r)}f$ over $\operatorname{osc}(r)\le R$ is
\begin{equation}
 \mathcal R(f,R)=
 \max_{\theta\in[0,1]}
 \frac{m_f+(\rho-1)U_f(\theta)}
      {1+(\rho-1)\theta}.
 \label{eq:robust-f}
\end{equation}
It suffices to evaluate $\theta$ at cumulative $q$-masses after sorting tokens by decreasing $f$. For position $t$, let $\mathcal R_t(f,R)$ denote the functional in~\eqref{eq:robust-f} evaluated with $q=q_t$ and $m_f=\mathbb E_{q_t}f$. In particular, define
\[
 \mathcal M_t(\lambda,R_t)=
 \mathcal R_t(e^{\lambda\ell_t},R_t).
\]
Consequently, $\mathcal M_t$ is the largest conditional moment of $e^{\lambda\ell_t}$.
For $\lambda<0$, this is the exact maximum conditional exponential moment under every admissible modified next-token law.
\end{theorem}


The exponential moment in Theorem~\ref{thm:mgf} connects one-step score changes to the detector's sequence-level miss rate. A miss occurs when the accumulated log-likelihood ratio falls below $\tau$. For $\lambda<0$, the quantity $e^{\lambda\ell}$ is larger for tokens with smaller evidence scores. A modification that shifts probability toward those tokens therefore raises this moment, making a miss harder to rule out.

At each prefix, Theorem~\ref{thm:mgf} computes the largest possible conditional average of $e^{\lambda\ell_t}$ among all logit changes allowed by the span budget $R_t$. To find it, sort the vocabulary tokens from largest to smallest $e^{\lambda\ell_t}$—equivalently, from smallest to largest $\ell_t$—and consider how much probability mass the modification can shift toward that ordering. The theorem shows that it is enough to check the cumulative probability masses at the token boundaries; no search over all possible residual vectors is needed.

The prefix changes as tokens are generated, so the next-step bound is evaluated at each new history. These conditional moment bounds multiply across the sequence; taking logarithms turns their product into the accumulated quantity
\[
A_N(\lambda)=\sum_{t=1}^N\log\mathcal M_t(\lambda,R_t).
\]
This step does not require the generated tokens to be independent: each bound is conditional on the history so far. The next theorem uses independent certification continuations to bound how often a future path's $A_N$ exceeds the largest value observed during certification, then combines that bound with a Chernoff argument to control the miss probability after attack selection.

\subsection{Calibration selection with independent certification}

Calibration may choose both the transformation and $\lambda$. We freeze the transformation, threshold, $\lambda$, and predictable span budgets, then evaluate the pathwise moment envelope on independent certification continuations. Their maximum supplies a distribution-free upper tolerance limit for future paths.

\begin{theorem}[Bounding Missed Detections on Fresh Samples After Attack Selection]\label{thm:holdout}
Fix $\lambda<0$, the threshold $\tau$, and the predictable budgets $R_t$. Let an attacker be chosen using arbitrary calibration data and then frozen. On each of $n$ independent certification continuations from that selected modified model, compute
\[
 A_N^{(i)}(\lambda)=
 \sum_{t=1}^N\log\mathcal M_t(\lambda,R_t(c,Y_{<t}^{(i)})),
 \qquad C_n=\max_{i\le n}A_N^{(i)}(\lambda).
\]
For confidence error $\delta_{\mathrm{cert}}\in(0,1)$, put $\epsilon_{n,\delta_{\mathrm{cert}}}=1-\delta_{\mathrm{cert}}^{1/n}$. With probability at least $1-\delta_{\mathrm{cert}}$ over the certification sample,
\begin{equation}
 \widetilde\beta_N(c,\tau)
 \le \min\{1,\exp(C_n-\lambda\tau)+\epsilon_{n,\delta_{\mathrm{cert}}}\}.
 \label{eq:holdout}
\end{equation}
The attack, threshold, $\lambda$, and budgets must be fixed before certification. If $J$ candidates are certified simultaneously, allocate candidate-specific confidence errors $\delta_j>0$ with $\sum_{j=1}^J\delta_j\le\delta_{\mathrm{cert}}$; the union bound then gives simultaneous coverage.
\end{theorem}


The bound separates a future miss into two cases. On paths where the accumulated envelope $A_N$ is no larger than the largest value $C_n$ seen in certification, the exponential-moment argument gives a miss-probability bound of $\exp(C_n-\lambda\tau)$. The remaining case is a path whose envelope exceeds $C_n$. A distribution-free tolerance bound controls how much probability can lie in that tail:
\[
\widetilde\beta_N(c,\tau)
\le
\exp(C_n-\lambda\tau)+\Pr_{\rm fresh}(A_N>C_n)
\le
\exp(C_n-\lambda\tau)+\epsilon_{n,\delta_{\rm cert}}
\]
with probability at least $1-\delta_{\rm cert}$ over the certification sample. Here $C_n$ is the largest envelope value among the $n$ certification paths, and $\epsilon_{n,\delta_{\rm cert}}=1-\delta_{\rm cert}^{1/n}$ is the distribution-free allowance for future paths above that observed maximum. The first term controls misses on paths whose envelope is covered by certification; the second accounts for paths beyond that coverage.

Calibration data may be used to choose the attack and $\lambda$, but those choices must be frozen before certification. The independent certification sample then gives a valid bound for a new continuation without reusing data for selection. When certifying several candidates, allocating confidence errors whose sum is at most $\delta_{\rm cert}$ gives simultaneous coverage by the union bound. The guarantee requires a valid predictable span budget on every relevant history. Section~5 reports the resulting bound for each candidate; in this study all eight bounds equal $1.0$, so the certificate is valid but numerically uninformative.

\section{Experiments}

\subsection{Models, data, and transformations}
We instantiate the OpenStamp verifier on Llama-2-7B and Mistral-7B-v0.3 using the released implementation and fixed selector matrices trained with OpenStamp's OpenWebText semantic-alignment pipeline. The watermark seed is fixed, the strength is \(\delta_{\mathrm{wm}}=1.0\), and the green-list fraction is \(\gamma=0.25\). Prompts are sampled from eligible documents in the local RealNewsLike validation corpus, stored in the C4 (Colossal Clean Crawled Corpus) validation data. We retain documents that support a 50-token prompt plus 128 continuation tokens under both tokenizers, leaving 11,697 eligible documents. For each model, we draw 500 contexts with replacement for each of calibration, certification, and test using prompt-selection seed; generation seeds are derived from a fixed base seed and each draw ID. The same prompt identities and seeds are used across the two models and all transformations.

The candidate family is the clean marked checkpoint, bitsandbytes INT8 quantization, 30\% per-layer unstructured magnitude pruning, and a 500-step OpenWebText LoRA update. LoRA uses the first 50,000 documents from the saved 500,000-document OpenWebText pool, shuffled randomly with a fixed seed. We follow OpenStamp's fine-tuning entry point with LoRA rank 16, scaling parameter 32, dropout 0.1, and adapters on the attention and multilayer perceptron (MLP) projections \texttt{q\_proj}, \texttt{k\_proj}, \texttt{v\_proj}, \texttt{o\_proj}, \texttt{gate\_proj}, \texttt{up\_proj}, and \texttt{down\_proj}; the watermark output head is held fixed. Training uses sequence length 256, effective batch size 16 (microbatch 2 with gradient accumulation 8), gradient checkpointing, bfloat16 (BF16), Adafactor, learning rate \(2\times10^{-5}\), and a cosine schedule. All conditions generate by multinomial sampling from the raw temperature-1 softmax for exactly 64 steps. The end-of-sequence (EOS) token remains sampleable, and generation uses no additional decoding warper. This matches the generation law used to compute the paired-model likelihood scores in Eq.~\ref{eq:llr}.

The clean marked condition uses all 500 draws in each split. Base-law null continuations are generated for the 500 test prompts per model and used to estimate the FPR. Attacked conditions use the first 250 draws from each split, giving 250 calibration, certification, and test continuations per attack. The study uses one watermark seed and one prompt-pool repetition. Reported intervals quantify within-run sampling uncertainty.

\subsection{Selection, certification, and metrics}
We fix the threshold at $\tau=\log 100$, which gives the base-law bound $e^{-\tau}=0.01$, and set familywise confidence error to $\delta_{\mathrm{cert}}=0.05$. For every model-candidate pair, calibration selects $\lambda$ from $\{-0.1,-0.25,-0.5,-1\}$ by minimizing $\max_i A_N^{(i)}(\lambda)-\lambda\tau$. It then selects the candidate with the largest calibration FNR among those with calibration perplexity ratio at most 1.10. The candidate and its $\lambda$ are frozen before certification and test data are used. We certify all $J=2\times4=8$ pairs with confidence error $\delta_{\mathrm{cert}}/J=0.00625$, giving simultaneous familywise coverage by the union bound in Theorem~\ref{thm:holdout}. The certification sample contains 250 continuations per attacked candidate and 500 per clean candidate; the corresponding tolerance terms are 0.0201 and 0.0101.

For each transformed checkpoint, we compare its logits with the original marked checkpoint on identical token prefixes and compute the exact predictable span budget \(R_t(c,y_{<t})=\operatorname{osc}(r_t)\). We report fixed-threshold TPR and FNR, base-law FPR, the perplexity (PPL) ratio, mean residual span, and mean local half-evidence radius. The radius is the value from Corollary~\ref{cor:breakdown} at target \(\eta_t=k_{1,t}/2\), averaged over evaluated test prefixes where it is defined. We report 95\% confidence intervals (CIs); the TPR and FPR intervals use the Wilson method, and the FNR interval is the complement of the TPR interval. Let NLL denote negative log-likelihood, and let $\overline{\mathrm{NLL}}_M$ be its mean per-token value for model $M$ on the reference continuations. The PPL ratio is \(\exp(\overline{\mathrm{NLL}}_{\rm transformed}-\overline{\mathrm{NLL}}_{\rm marked})\), computed by teacher-forcing both models on the same 64-token reference continuation; a value of 1 means equal average likelihood on that text. Figure~\ref{fig:envelope-slack} reports prompt-bootstrap CIs for cumulative local-envelope gaps.

For the directional diagnostic, define \(\widetilde\mu_t=\mathbb E_{\widetilde q_t}[\ell_t]\). We compute the conditional mean drift \(\widetilde\mu_t-k_{1,t}\) and its first-order covariance term \(\operatorname{Cov}_{q_t}(d_t,r_t)\) at each test prefix, using full-vocabulary logits from the base, marked, and transformed checkpoints. We aggregate these quantities by continuation and report 95\% percentile-bootstrap CIs over continuations. This reuses the test continuations and adds no generation or model training; it is a retrospective check of the local expansion, not a separate TPR estimate or a prospective attack-selection rule.

\subsection{Detection and perplexity}
Table~\ref{tab:main-experiments} gives fixed-threshold detection and perplexity results. INT8 yields TPRs of 94.8\% for Llama-2-7B and 86.0\% for Mistral-7B-v0.3; 30\% pruning yields 90.8\% and 85.6\%, respectively. These rates are close to the clean marked checkpoints (94.8\% and 85.4\%). LoRA lowers TPR to 28.8\% for Llama and 11.6\% for Mistral, corresponding to FNRs of 71.2\% and 88.4\%. The formal fresh-holdout bounds are discussed separately in Section~5.6.

Each model has zero false positives among 500 base-law test continuations; the 95\% Wilson upper endpoint is 0.76\%, rounded to 0.8\% in Table~\ref{tab:main-experiments}. The likelihood ratio on fixed reference text increases under pruning (1.185 for Llama and 1.065 for Mistral) and is slightly below one after LoRA (0.911 and 0.949). These ratios describe teacher-forced likelihood on the evaluation continuations. LoRA's test areas under the receiver operating characteristic curve (AUROCs) remain 0.975 and 0.930, showing that the score still ranks most modified-model samples above base-law samples even though recall at the fixed threshold \(\log 100\) is low.

\begin{table*}[t]
\caption{Fixed-threshold results for the clean marked checkpoints and three post-release transformations. TPR and FPR are percentages with 95\% Wilson intervals. The final column reports observed test FNR ($100\%-\mathrm{TPR}$), with its interval obtained by complementing the TPR interval. Each model's FPR uses 500 base-law continuations; clean TPR uses 500 continuations and each attacked TPR uses 250. Mean span and mean $R^*_{1/2}$ average the per-prefix residual span and local half-evidence radius over test continuations. The radius is the span at which the least-favorable residual halves that prefix's conditional mean LLR. PPL ratio is the transformed-to-marked perplexity ratio on the same reference continuations; 1.0 indicates equal mean NLL.}
\label{tab:main-experiments}
\centering
\small
\resizebox{\textwidth}{!}{%
\begin{tabular}{llrrrrrr}
\toprule
Model & Modification & Mean span & Mean $R^*_{1/2}$ & TPR (95\% CI) & FPR (95\% CI) & PPL ratio & Observed FNR (95\% CI)\\
\midrule
Llama-2-7B & Unmodified marked & 0.00 & 0.44 & 94.8 [92.5, 96.4] & 0.0 [0.0, 0.8] & 1.000 & 5.2 [3.6, 7.5]\\
Llama-2-7B & INT8 & 1.78 & 0.44 & 94.8 [91.3, 96.9] & 0.0 [0.0, 0.8] & 1.006 & 5.2 [3.1, 8.7]\\
Llama-2-7B & 30\% pruning & 7.91 & 0.43 & 90.8 [86.6, 93.8] & 0.0 [0.0, 0.8] & 1.185 & 9.2 [6.2, 13.4]\\
Llama-2-7B & OpenWebText LoRA (500 steps) & 10.42 & 0.44 & 28.8 [23.5, 34.7] & 0.0 [0.0, 0.8] & 0.911 & 71.2 [65.3, 76.5]\\
Mistral-7B-v0.3 & Unmodified marked & 0.00 & 0.36 & 85.4 [82.0, 88.2] & 0.0 [0.0, 0.8] & 1.000 & 14.6 [11.8, 18.0]\\
Mistral-7B-v0.3 & INT8 & 1.07 & 0.37 & 86.0 [81.2, 89.8] & 0.0 [0.0, 0.8] & 1.001 & 14.0 [10.2, 18.8]\\
Mistral-7B-v0.3 & 30\% pruning & 4.42 & 0.37 & 85.6 [80.7, 89.4] & 0.0 [0.0, 0.8] & 1.065 & 14.4 [10.6, 19.3]\\
Mistral-7B-v0.3 & OpenWebText LoRA (500 steps) & 9.03 & 0.36 & 11.6 [8.2, 16.2] & 0.0 [0.0, 0.8] & 0.949 & 88.4 [83.8, 91.8]\\
\bottomrule
\end{tabular}%
}
\end{table*}

\subsection{Numerical checks of the evidence envelopes}
We evaluate the finite-breakpoint formulas at the visited prefixes for all eight model-condition pairs. These checks test the exact mean lower envelope and its attainable witness in Theorem~\ref{thm:envelope}, Eq.~\ref{eq:mean-envelope}, the span-only bound in Eq.~\ref{eq:tanh}, and the exponential-moment oracle and its witness in Theorem~\ref{thm:mgf}, Eq.~\ref{eq:robust-f}. All four numerical checks pass at tolerance $10^{-6}$. More precisely, modified-law conditional means lie above the exact lower envelope, the exact envelope lies above the summed $\tanh$ lower bound, and both constructed witnesses attain their corresponding optima. The largest mean-witness gap is $1.71\times10^{-13}$; the largest mean absolute moment-witness gap is $1.45\times10^{-13}$. The two minimum slacks are zero up to numerical precision, as expected for the clean condition, where residual span is zero. These checks confirm that the finite-vocabulary calculations implement the formulas on the evaluated prefixes.

\begin{figure*}[t]
\centering
\includegraphics[width=.84\textwidth]{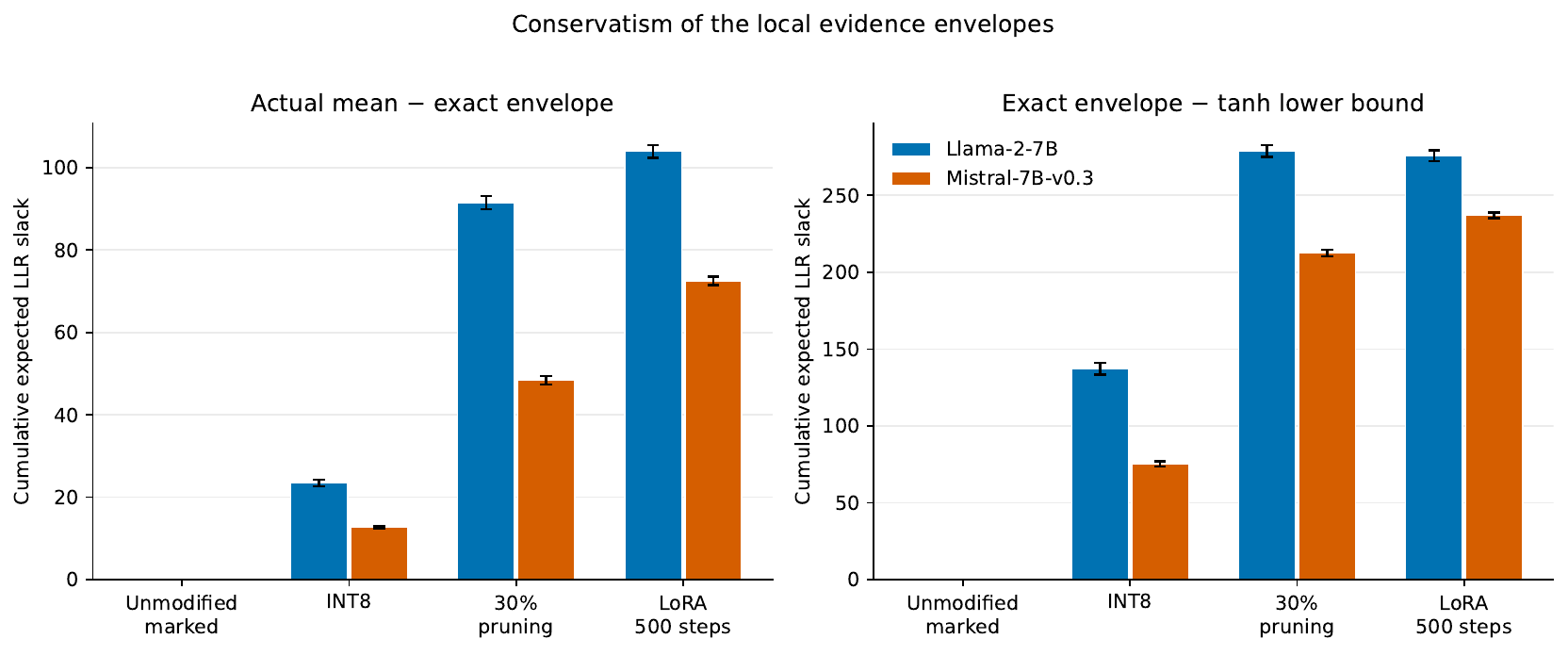}
\caption{Each bar is the prompt-bootstrap mean of the 64-step cumulative gap on test continuations; whiskers are 95\% percentile-bootstrap intervals over prompts. The left panel compares modified-law conditional expected LLR with the exact span-constrained lower envelope in Theorem~\ref{thm:envelope}, Eq.~\ref{eq:mean-envelope}; the right compares that envelope with the summed $\tanh$ lower bound in Eq.~\ref{eq:tanh}. Both gaps are zero for the clean checkpoint because its residual span is zero.}
\Description{Two panels compare cumulative LLR slack for Llama-2-7B and Mistral-7B-v0.3 across unmodified, INT8, pruned, and LoRA models. Both gaps grow under modification; the exact envelope is tighter than the tanh relaxation.}
\label{fig:envelope-slack}
\end{figure*}

Figure~\ref{fig:envelope-slack} compares the local bounds along the test continuations. For each prompt, we sum the 64 prefixwise gaps and then average across prompts; whiskers are 95\% percentile-bootstrap intervals over prompts. The left panel is the modified-law conditional expected LLR minus the exact worst-case lower envelope in Theorem~\ref{thm:envelope}, Eq.~\ref{eq:mean-envelope}. Its mean is 23.47, 91.54, and 103.95 LLR units for Llama under INT8, pruning, and LoRA, and 12.65, 48.35, and 72.55 for Mistral. The right panel is the exact envelope minus the summed span-only $\tanh$ lower bound from Eq.~\ref{eq:tanh}, with means 137.33, 278.79, and 275.69 for Llama and 75.15, 212.39, and 237.05 for Mistral. The first gap shows how much the actual residual direction leaves above the worst direction allowed by its span. The second shows the extra slack from a span-only bound that omits token-level probabilities and scores. The distribution-aware envelope is tighter on these checkpoints. Both curves summarize local worst-case diagnostics.

Across modified checkpoints, mean residual spans range from 1.07 to 10.42, while mean local half-evidence radii range from 0.36 to 0.44. These radii use Corollary~\ref{cor:breakdown}, Eq.~\ref{eq:radius}, with target $\eta=k_1/2$; each is the per-prefix threshold for the score-ordered least-favorable residual. Actual quantization, pruning, and LoRA residuals follow their own directions, so their TPRs are interpreted together with the envelope gaps in Figure~\ref{fig:envelope-slack}.

\subsection{Directional evidence drift}
\begin{figure*}[t]
\centering
\includegraphics[width=.84\textwidth]{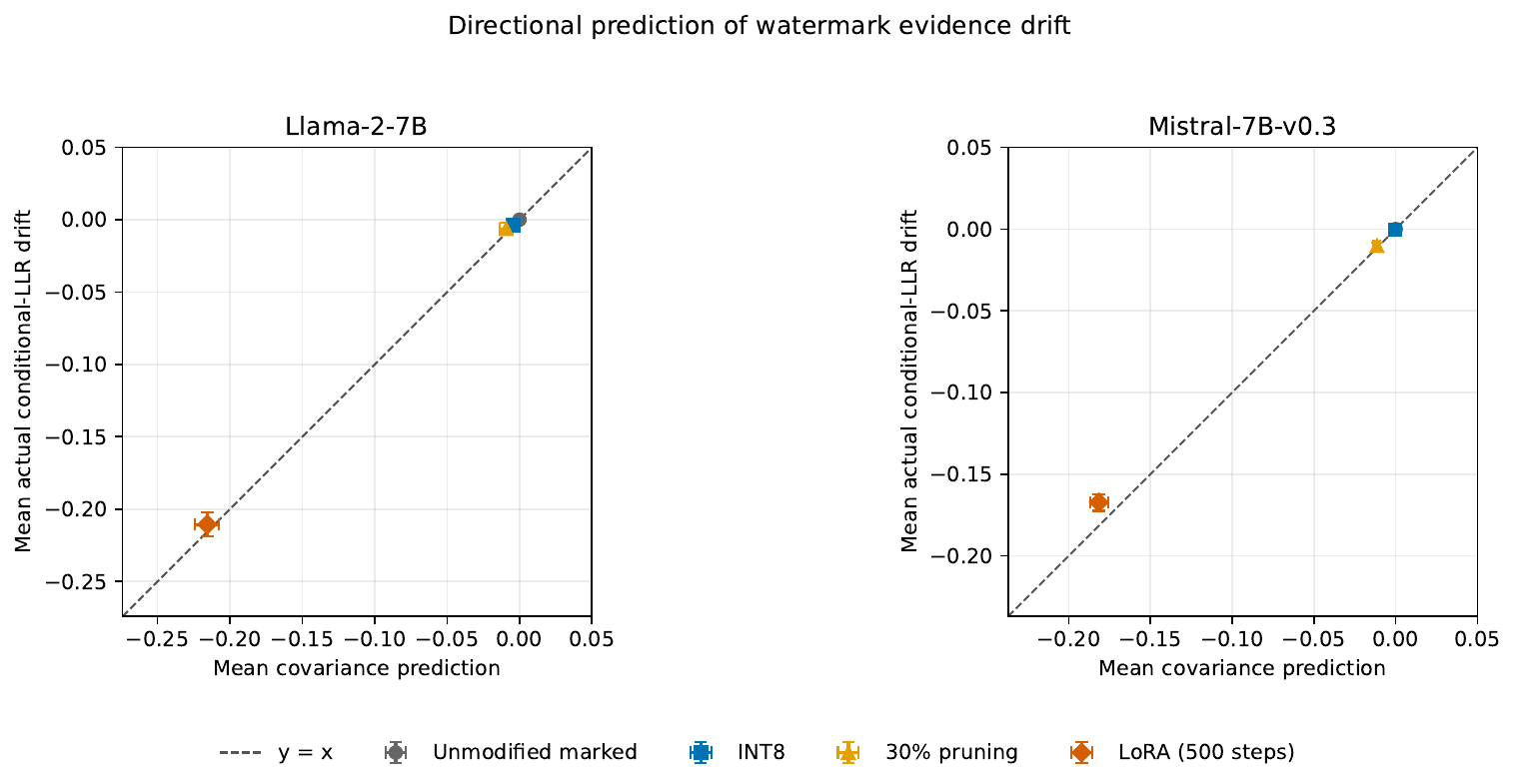}
\caption{Mean exact conditional LLR drift, defined in Eq.~\ref{eq:drift}, versus the first-order covariance prediction from Corollary~\ref{cor:directional}, Eq.~\ref{eq:directional-expansion}, on test continuations. Each point is one model-condition pair; the dashed diagonal marks equality. Whiskers are 95\% percentile-bootstrap intervals over continuations. The second-order remainder bound in Eq.~\ref{eq:directional-expansion} has zero observed prefixwise violations across 160,000 prefixes, within floating-point tolerance.}
\Description{Two panels compare mean actual conditional LLR drift to its covariance prediction for Llama-2-7B and Mistral-7B-v0.3. INT8 and pruning have small drifts, LoRA has larger negative drift, and the plotted points lie near the equality line.}
\label{fig:directional-drift}
\end{figure*}

The span envelope allows every logit residual with the specified oscillation, so it describes a worst-case probability shift (Theorem~\ref{thm:envelope}, Eqs.~\ref{eq:mean-envelope}--\ref{eq:tanh}). For a particular residual, Corollary~\ref{cor:directional}, Eq.~\ref{eq:directional-expansion}, gives the first-order mean-evidence change through its covariance with the watermark scores. Figure~\ref{fig:directional-drift} compares this covariance prediction with the exact conditional mean drift defined by Eq.~\ref{eq:drift} over the test prefixes.

The point estimates have the same direction in all six modified model-condition groups. INT8 and pruning produce small mean drifts, from $-3.5\times10^{-4}$ to $-1.03\times10^{-2}$ LLR units per token. LoRA produces larger decreases: for Llama, the measured and predicted means are $-0.211$ and $-0.216$; for Mistral, they are $-0.167$ and $-0.182$. These comparisons evaluate the covariance term and explicit remainder in Corollary~\ref{cor:directional}, Eq.~\ref{eq:directional-expansion}. The held-out Mistral INT8 drift is close to zero, and its bootstrap interval includes zero. Across all eight groups, including the two clean checkpoints, the second-order remainder bound in Eq.~\ref{eq:directional-expansion} has zero observed prefixwise violations across 160,000 prefixes, up to the stated floating-point tolerance.

The remainder bound in Corollary~\ref{cor:directional}, Eq.~\ref{eq:directional-expansion}, is conservative for these checkpoints: for Llama LoRA, the mean absolute remainder is $0.046$ LLR units while the mean allowance is $91.43$; for Mistral LoRA, the corresponding values are $0.037$ and $59.26$. The covariance term therefore gives a useful empirical account of the average local direction, while the theorem's span-based remainder remains a valid but loose worst-case allowance. This diagnostic helps interpret the larger LoRA detection losses in Table~\ref{tab:main-experiments}; it does not itself predict the fixed-threshold TPR or provide an additional population-risk guarantee.

\subsection{Calibration selection and fresh-holdout results}
The calibration rule selects the 500-step OpenWebText LoRA candidate for both models. Its calibration FNRs are 0.728 for Llama and 0.872 for Mistral, and its PPL ratios are 0.905 and 0.944, within the prespecified 1.10 eligibility limit. On test data, the selected candidates have FNRs of 0.712 and 0.884. Calibration chooses $\lambda=-0.1$ for all eight model-candidate pairs. The selected transformation and tuning parameters are frozen before certification, as required by Theorem~\ref{thm:holdout}; the test FNRs are empirical estimates, distinct from the theorem's confidence bound.

The familywise 95\% holdout upper bound is 1.000 for each of the eight candidates. This evaluates the fresh-holdout guarantee in Theorem~\ref{thm:holdout}, Eq.~\ref{eq:holdout}, using the pathwise moment envelope from Theorem~\ref{thm:mgf}, Eq.~\ref{eq:robust-f}. The order-statistic allowance is 0.0201 for attacked candidates and 0.0101 for clean candidates; the moment term makes the combined bound clip at one. Table~\ref{tab:main-experiments} reports the corresponding observed test FNRs, which are finite-sample rates rather than population bounds. Calibration selects LoRA as the highest-miss eligible candidate in both models, and the independent holdout procedure preserves simultaneous coverage after that selection, as guaranteed by Theorem~\ref{thm:holdout}. For this run, the moment term prevents the certificate from giving a useful numerical miss-risk bound.

\section{Conclusion}

We give an exact account of how a bounded logit residual can change the expected evidence of a fixed low-rank watermark, then carry the one-step analysis into a calibration-selected, fresh-holdout miss-risk bound. Numerical witnesses match the finite-vocabulary formulas, and the first-order covariance term tracks the direction of mean evidence drift on 160,000 test prefixes. INT8 and pruning retain test TPRs near the clean checkpoints; 500-step LoRA lowers TPR to 28.8\% for Llama and 11.6\% for Mistral. The holdout bound is 1.0 for all eight candidates, so the current sample and moment envelope do not yield a useful numerical population miss-risk guarantee. Larger certification sets, additional seeds, and more modification types are needed to assess when the certificate becomes informative.

\appendix
\section{Proofs}
\setlength{\abovedisplayskip}{3pt plus 1pt minus 1pt}
\setlength{\belowdisplayskip}{3pt plus 1pt minus 1pt}
\setlength{\abovedisplayshortskip}{2pt plus 1pt minus 1pt}
\setlength{\belowdisplayshortskip}{2pt plus 1pt minus 1pt}

\begin{proof}[Proof of Theorem~\ref{thm:envelope}]
First consider the interpolation identity. For $u\in[0,1]$, define
\[
\begin{aligned}
 Z_u&=\sum_v e^{z(v)+d(v)+ur(v)},\\
 q_u(v)&=e^{z(v)+d(v)+ur(v)}/Z_u.
\end{aligned}
\]
Differentiating gives
\[
 q'_u(v)=q_u(v)\bigl(r(v)-\mathbb E_{q_u}r\bigr).
\]
Thus
\[
 \frac{d}{du}\mathbb E_{q_u}d
 =\sum_vq_u(v)d(v)r(v)
  -\mathbb E_{q_u}d\,\mathbb E_{q_u}r
 =d^\top(\operatorname{diag}(q_u)-q_uq_u^\top)r.
\]
Integrating from $0$ to $1$ gives $\mathbb E_{\widetilde q}d-\mathbb E_qd$. Since $\ell(v)=d(v)-\log\sum_up(u)e^{d(u)}$ differs from $d$ by a token-independent constant, the same expectation difference holds for $\ell$, proving~\eqref{eq:drift}.

For the envelope, let $w(v)=\widetilde q(v)/q(v)$. Softmax normalization gives $w=e^r/\mathbb E_qe^r$, so $\mathbb E_qw=1$ and $\max w/\min w=e^{\operatorname{osc}(r)}\le\rho$. Conversely, any positive normalized $w$ with this ratio bound is induced by $r=\log w$. For $\rho>1$, put $a=\min_vw(v)$ and $u_v=(w(v)/a-1)/(\rho-1)$. Then $0\le u_v\le1$. With $\theta=\mathbb E_qu$, normalization yields
\begin{equation}
 w(v)=\frac{1+(\rho-1)u_v}{1+(\rho-1)\theta}.
 \label{eq:proof-density-ratio}
\end{equation}
Every such $u$ also defines an admissible $w$, so this parameterization is exact. For $\rho=1$, normalization forces $w\equiv1$.

Fix $\theta$. In the nondecreasing score order, set $x_i=q(v_i)u_{v_i}$. The constraints become $0\le x_i\le q(v_i)$ and $\sum_i x_i=\theta$. If a higher-score token has $x_b>0$ while a lower-score token has unused capacity, moving $\varepsilon=\min\{x_b,q(v_a)-x_a\}$ units from $b$ to $a$ changes the objective $\sum_i x_i\ell(v_i)$ by $\varepsilon(\ell(v_a)-\ell(v_b))\le0$. Hence a minimizer fills lowest-score tokens first. For $Q_j\le\theta\le Q_{j+1}$ the minimum allocated score is $C_j+(\theta-Q_j)\ell(v_{j+1})$, and the mean is
\[
 F_j(\theta)=
 \frac{k_1+(\rho-1)[C_j+(\theta-Q_j)\ell(v_{j+1})]}
      {1+(\rho-1)\theta}.
\]
Differentiating the quotient gives
\[
 F'_j(\theta)=
 \frac{(\rho-1)\{\ell(v_{j+1})[1+(\rho-1)Q_j]-k_1-(\rho-1)C_j\}}
      {[1+(\rho-1)\theta]^2}.
\]
Its numerator is constant on this interval, so $F_j$ is monotone or constant there and its minimum is at an endpoint $Q_j$. Substitution gives~\eqref{eq:mean-envelope}. At an endpoint $u$ is $1$ on the first $j$ tokens and $0$ elsewhere; equation~\eqref{eq:proof-density-ratio} then gives a two-level $w$, hence a two-level residual $r=\log w$.

For distributions $\mu$ and $\nu$ on $\mathcal V$, define their total-variation distance by $\operatorname{TV}(\mu,\nu)=\frac12\sum_{v\in\mathcal V}|\mu(v)-\nu(v)|$. For a real number $x$, write $x_+=\max\{x,0\}$. For the span-only bound, the case $R=0$ gives $w\equiv1$ and is immediate. For $R>0$, write $w=\widetilde q/q$, $a=\min w$, and $b=\max w$. If $a=b$, total variation is zero; otherwise $a\le1\le b$. Since $\mathbb E_qw=1$,
\[
 \operatorname{TV}(\widetilde q,q)=\mathbb E_q(w-1)_+
 \le\frac{(b-1)(1-a)}{b-a}.
\]
The inequality follows pointwise from $(x-1)_+\le (b-1)(x-a)/(b-a)$ for $x\in[a,b]$. For fixed $a$ its right side increases with $b$, so substituting $b\le\rho a$ gives
\[
 \operatorname{TV}(\widetilde q,q)
 \le\frac{\rho+1-\rho a-a^{-1}}{\rho-1}.
\]
The derivative of the right side in $a$ is $(a^{-2}-\rho)/(\rho-1)$; its maximum on $[1/\rho,1]$ is at $a=1/\sqrt\rho$, with value $(\sqrt\rho-1)/(\sqrt\rho+1)=\tanh(R/4)$. Equality is attained by a two-point $w$ taking $1/\sqrt\rho$ and $\sqrt\rho$, with probability $1/(\sqrt\rho+1)$ on the larger value so that $\mathbb E_qw=1$.

For any $g$, $|\mathbb E_\mu g-\mathbb E_\nu g|\le\operatorname{osc}(g)\operatorname{TV}(\mu,\nu)$: the positive and negative parts of the signed measure $\mu-\nu$ each have mass $\operatorname{TV}(\mu,\nu)$, and $g$ ranges over an interval of length $\operatorname{osc}(g)$. Apply this with $g=d$, $\mu=\widetilde q$, and $\nu=q$. The two-point construction with the score extrema aligned attains the bound. Finally, $x-\tanh x$ has derivative $\tanh^2x\ge0$ and is zero at $x=0$, proving $\tanh x\le x$ and the linear bound in~\eqref{eq:tanh}.
\end{proof}

\begin{proof}[Proof of Corollary~\ref{cor:directional}]
For $u\in[0,1]$, define
\[
 q_u(v)=\operatorname{softmax}(z+d+ur)(v),
 \qquad \phi(u)=\mathbb E_{q_u}[d(X)].
\]
The endpoints are $q_0=q$ and $q_1=\widetilde q$. Differentiating the softmax normalizer gives
\[
 q'_u(v)=q_u(v)\bigl(r(v)-\mathbb E_{q_u}r(X)\bigr).
\]
Therefore
\begin{align*}
 \phi'(u)
 &=\sum_vd(v)q'_u(v)\\
 &=\mathbb E_{q_u}[d(X)r(X)]
   -\mathbb E_{q_u}[d(X)]\mathbb E_{q_u}[r(X)]\\
 &=\operatorname{Cov}_{q_u}(d(X),r(X)).
\end{align*}
Differentiating once more yields
\begin{equation}
 \phi''(u)=\mathbb E_{q_u}\!\left[
  (d(X)-\mathbb E_{q_u}d(X))
  (r(X)-\mathbb E_{q_u}r(X))^2
 \right].
 \label{eq:proof-directional-second}
\end{equation}
To verify this expression explicitly, write $a=\mathbb E_{q_u}[d(X)]$, $b=\mathbb E_{q_u}[r(X)]$, and $c=\mathbb E_{q_u}[d(X)r(X)]$. Then $\phi'=c-ab$, $c'=\mathbb E_{q_u}[d(X)r(X)^2]-bc$, $a'=c-ab$, and $b'=\mathbb E_{q_u}[r(X)^2]-b^2$. Substitution into $\phi''=c'-a'b-ab'$ gives
\[
 \phi''(u)=\mathbb E_{q_u}[d(X)r(X)^2]-2b\mathbb E_{q_u}[d(X)r(X)]
         -a\mathbb E_{q_u}[r(X)^2]+2ab^2,
\]
which equals the centered expression in~\eqref{eq:proof-directional-second}.

Let $A_d=\operatorname{osc}(d)$ and $A_r=\operatorname{osc}(r)$. Since $d(X)$ lies in an interval of length $A_d$, its centered value has absolute value at most $A_d$. For a real random variable $Z$, define $\operatorname{Var}(Z)=\mathbb E[(Z-\mathbb E Z)^2]$; in particular,
\[
 \operatorname{Var}_{q_u}(r(X))=\mathbb E_{q_u}[(r(X)-\mathbb E_{q_u}r(X))^2].
\]
Therefore
\[
 |\phi''(u)|\le A_d\operatorname{Var}_{q_u}(r(X)).
\]
For completeness, if a random variable \(Z\) lies in \([m,M]\) and has mean \(\mu\), then \((Z-m)(M-Z)\ge0\). Expanding and taking expectations gives \(\mathbb E[Z^2]\le(m+M)\mu-mM\), and therefore
\[
 \operatorname{Var}(Z)\le(M-\mu)(\mu-m)
 \le\frac{(M-m)^2}{4}.
\]
Applying this to \(Z=r(X)\) under \(q_u\) gives \(|\phi''(u)|\le A_dA_r^2/4\) for every \(u\).

Taylor's formula with integral remainder now gives
\[
 \phi(1)=\phi(0)+\phi'(0)+\int_0^1(1-u)\phi''(u)\,du,
\]
so
\[
 \left|\phi(1)-\phi(0)-\phi'(0)\right|
 \le\frac{A_dA_r^2}{4}\int_0^1(1-u)\,du
 =\frac18A_dA_r^2.
\]
Finally, $\ell(v)=d(v)-\log\sum_up(u)e^{d(u)}$ differs from $d(v)$ by a constant. Thus $\phi(1)-\phi(0)=\widetilde\mu-k_1$, while $\phi'(0)=\operatorname{Cov}_{q}(d(X),r(X))$. This proves~\eqref{eq:directional-expansion}.
\end{proof}

\begin{proof}[Proof of Corollary~\ref{cor:breakdown}]
For a candidate prefix size $j\ge1$, the envelope endpoint has mean
\[
 F_j(\rho)=\frac{k_1+(\rho-1)C_j}{1+(\rho-1)Q_j}.
\]
Because the denominator is positive, $F_j(\rho)\le\eta$ is equivalent to
\[
 k_1-\eta\le(\rho-1)(\eta Q_j-C_j).
\]
The left side is positive, so this prefix is feasible exactly when $\eta Q_j-C_j>0$, or $C_j/Q_j<\eta$. Its smallest feasible ratio is $\rho_j=1+(k_1-\eta)/(\eta Q_j-C_j)$. Minimizing $R=\log\rho_j$ over feasible $j$ gives~\eqref{eq:radius}. At the minimizing prefix equality holds, and the two-level residual from Theorem~\ref{thm:envelope} attains mean $\eta$. The condition $\eta>\min_v\ell(v)$ guarantees at least one feasible prefix, namely the first token.

Write \(R^\star=R^\star(\eta)\), and let \(j^\star\) attain the minimum in~\eqref{eq:radius}. Let \(\mathbf 1\in\mathbb R^V\) denote the all-ones vector. Let \(r^\star\) equal \(R^\star\) on the selected \(j^\star\) lowest-scoring tokens and zero elsewhere. Its centered version \(\bar r^\star=r^\star-(j^\star R^\star/V)\mathbf 1\) has the same softmax effect. Its squared norm is
\[
\begin{aligned}
 \|\bar r^\star\|_2^2
 &=j^\star\left(R^\star\frac{V-j^\star}{V}\right)^2
   +(V-j^\star)\left(R^\star\frac{j^\star}{V}\right)^2\\
 &=(R^\star)^2\frac{j^\star(V-j^\star)}{V}.
\end{aligned}
\]
For $h\ne0$, define $A^\star=\bar r^\star h^\top/\|h\|_2^2$. Then $A^\star h=\bar r^\star$ and its rank is at most one. Since $\|uv^\top\|_F=\|u\|_2\|v\|_2$,
\[
 \|A^\star\|_F=\frac{R^\star}{\|h\|_2}\sqrt{\frac{j^\star(V-j^\star)}{V}}.
\]
Any residual inducing the same conditional distribution as $\bar r^\star$ has the form $\bar r^\star+c\mathbf 1$ for some $c\in\mathbb R$. Since $\mathbf 1^\top\bar r^\star=0$,
\[
 \|\bar r^\star+c\mathbf 1\|_2^2
 =\|\bar r^\star\|_2^2+Vc^2,
\]
so centering minimizes the norm among all residuals with that softmax distribution. For any target vector $x$ and any matrix $A$ satisfying $Ah=x$, writing the rows of $A$ as $a_v^\top$, Cauchy--Schwarz gives $\|a_v\|_2\ge|x(v)|/\|h\|_2$. Hence the least possible Frobenius norm for that target is $\|x\|_2/\|h\|_2$, attained by $A=xh^\top/\|h\|_2^2$. Taking $x=\bar r^\star+c\mathbf 1$ and minimizing over $c$ shows that $A^\star$ is minimum norm among all updates inducing the stated conditional distribution.
\end{proof}

\begin{proof}[Proof of Theorem~\ref{thm:mgf}]
Use~\eqref{eq:proof-density-ratio} and fix $\theta=\mathbb E_qu$. Then
\[
 \mathbb E_{\widetilde q}f
 =\frac{m_f+(\rho-1)\sum_vq(v)u_vf(v)}
       {1+(\rho-1)\theta}.
\]
At fixed $\theta$, the denominator is fixed. Writing $x_v=q(v)u_v$, moving allocated mass from a lower-$f$ token with $x_v>0$ to a higher-$f$ token with $x_v<q(v)$ increases the numerator and preserves feasibility. Therefore the maximum fills tokens in decreasing order of $f$, giving $U_f(\theta)$. In this order define $Q_j^f=\sum_{i\le j}q(v_i)$ and $B_j^f=\sum_{i\le j}q(v_i)f(v_i)$. For $Q_j^f\le\theta\le Q_{j+1}^f$, $U_f(\theta)=B_j^f+(\theta-Q_j^f)f(v_{j+1})$. Differentiating the ratio in~\eqref{eq:robust-f} gives
\[
 \frac{(\rho-1)\{f(v_{j+1})[1+(\rho-1)Q_j^f]-m_f-(\rho-1)B_j^f\}}
      {[1+(\rho-1)\theta]^2}.
\]
Its numerator is constant on the interval. The ratio is therefore monotone or constant, so its maximum is at an endpoint. Evaluating the endpoints proves~\eqref{eq:robust-f}; setting $f=e^{\lambda\ell}$ gives the claimed moment oracle.
\end{proof}

\begin{proof}[Proof of Theorem~\ref{thm:holdout}]
Condition on the calibration data and on the selected attacker, $\lambda$, $\tau$, and predictable budgets, all of which are frozen before certification. Fix any realization of the certification sample; this makes $C_n$ fixed. Consider an independent fresh continuation from the selected modified model. Let $\mathcal F_t$ be its token history, and set $A_t=\sum_{s=1}^t\log\mathcal M_s(\lambda,R_s)$, $Z_t=\exp(\lambda\sum_{s=1}^t\ell_s-A_t)$, and $Z_0=1$. Since $\mathcal M_t$ is prefix-measurable, Theorem~\ref{thm:mgf} gives
\[
 \mathbb E[Z_t\mid\mathcal F_{t-1}]
 =Z_{t-1}\frac{\mathbb E[e^{\lambda\ell_t}\mid\mathcal F_{t-1}]}{\mathcal M_t}
 \le Z_{t-1},
\]
so $Z_t$ is a nonnegative supermartingale and $\mathbb E Z_N\le1$. On $\{\Lambda_N<\tau,A_N\le C_n\}$, $\lambda<0$ implies $Z_N>e^{\lambda\tau-C_n}$. Markov's inequality therefore gives, conditional on this certification realization,
\[
 \Pr_{\rm fresh}(\Lambda_N<\tau,A_N\le C_n)
 \le e^{C_n-\lambda\tau}.
\]

It remains to bound the probability mass of fresh paths for which $A_N>C_n$. Under the fixed calibration choices, let $X=A_N$ for a fresh continuation, with CDF $F_X$, and let $X_1,\ldots,X_n$ be the i.i.d. certification values, so $C_n=\max_iX_i$. If $\Pr(X>C_n)>\epsilon$, then $F_X(C_n)<1-\epsilon$. The set $B=\{x:F_X(x)<1-\epsilon\}$ is lower, so every $X_i\le C_n$ lies in $B$. Consequently,
\[
 \Pr(C_n\in B)\le\Pr(X_i\in B\text{ for all }i)
 =\Pr(X\in B)^n\le(1-\epsilon)^n,
\]
including when $X$ has atoms. Taking $\epsilon=1-\delta_{\mathrm{cert}}^{1/n}$ shows that, with probability at least $1-\delta_{\mathrm{cert}}$ over the certification sample, $\Pr_{\rm fresh}(A_N>C_n)\le\epsilon$. On this event, split the miss event according to whether $A_N\le C_n$ or $A_N>C_n$ and apply the preceding Markov bound to obtain~\eqref{eq:holdout}. Allocating errors $\delta_j$ with $\sum_j\delta_j\le\delta_{\mathrm{cert}}$ and applying the union bound proves simultaneous coverage.

\end{proof}

\normalsize
\section*{GenAI Usage Disclosure}
Generative AI tools assisted with literature organization and drafting; the authors verified the sources, proofs, and reported results.

\bibliographystyle{ACM-Reference-Format}
\bibliography{references}

@inproceedings{kirchenbauer2023watermark,
  title={A watermark for large language models},
  author={Kirchenbauer, John and Geiping, Jonas and Wen, Yuxin and Katz, Jonathan and Miers, Ian and Goldstein, Tom},
  booktitle={International conference on machine learning},
  pages={17061--17084},
  year={2023},
  organization={PMLR}
}

@inproceedings{hu2024unbiased,
  title={Unbiased watermark for large language models},
  author={Hu, Zhengmian and Chen, Lichang and Wu, Xidong and Wu, Yihan and Zhang, Hongyang and Huang, Heng},
  booktitle={International Conference on Learning Representations},
  volume={2024},
  pages={45408--45436},
  year={2024}
}

@inproceedings{bansal2022certified,
  title={Certified neural network watermarks with randomized smoothing},
  author={Bansal, Arpit and Chiang, Ping-yeh and Curry, Michael J and Jain, Rajiv and Wigington, Curtis and Manjunatha, Varun and Dickerson, John P and Goldstein, Tom},
  booktitle={International Conference on Machine Learning},
  pages={1450--1465},
  year={2022},
  organization={PMLR}
}

@article{cohen2023hilbert,
  title={Hyperbolic contractivity and the Hilbert metric on probability measures},
  author={Cohen, Samuel N and Fausti, Eliana},
  journal={arXiv preprint arXiv:2309.02413},
  year={2023}
}

@article{benavoli2013densityratio,
  title={Density-ratio robustness in dynamic state estimation},
  author={Benavoli, Alessio and Zaffalon, Marco},
  journal={Mechanical Systems and Signal Processing},
  volume={37},
  number={1-2},
  pages={54--75},
  year={2013},
  publisher={Elsevier}
}

@article{zhao2024provable,
  title={Provable robust watermarking for ai-generated text},
  author={Zhao, Xuandong and Ananth, Prabhanjan and Li, Lei and Wang, Yu-Xiang},
  journal={arXiv preprint arXiv:2306.17439},
  year={2023}
}

@inproceedings{pratihar2026catch22,
  title={Catch-22: On the Fundamental Tradeoff Between Detectability and Robustness in LLM Watermarking},
  author={Pratihar, Kuheli and Mukhopadhyay, Debdeep},
  booktitle={Forty-third International Conference on Machine Learning}
}

@article{cai2026statistical,
  title={Towards better statistical understanding of watermarking llms},
  author={Cai, Zhongze and Liu, Shang and Wang, Hanzhao and Zhong, Huaiyang and Li, Xiaocheng},
  journal={Journal of the American Statistical Association},
  volume={121},
  number={553},
  pages={247--258},
  year={2026},
  publisher={Taylor \& Francis}
}

@article{he2025dawa,
  title={Theoretically grounded framework for llm watermarking: A distribution-adaptive approach},
  author={He, Haiyun and Liu, Yepeng and Wang, Ziqiao and Mao, Yongyi and Bu, Yuheng},
  journal={Advances in Neural Information Processing Systems},
  volume={38},
  pages={103426--103462},
  year={2026}
}

@article{howard2020chernoff,
  author  = {Howard, Steven R. and Ramdas, Aaditya and McAuliffe, Jon and Sekhon, Jasjeet},
  title   = {Time-Uniform Chernoff Bounds via Nonnegative Supermartingales},
  journal = {Probability Surveys},
  volume  = {17},
  pages   = {257--317},
  year    = {2020},
  doi     = {10.1214/18-PS321},
  url     = {https://arxiv.org/abs/1808.03204}
}

@article{wilks1941tolerance,
  author  = {Wilks, Samuel S.},
  title   = {Determination of Sample Sizes for Setting Tolerance Limits},
  journal = {The Annals of Mathematical Statistics},
  volume  = {12},
  number  = {1},
  pages   = {91--96},
  year    = {1941},
  doi     = {10.1214/aoms/1177731788},
  url     = {https://doi.org/10.1214/aoms/1177731788}
}

@inproceedings{huang2026rlcracker,
  author    = {Huang, Hanbo and Zhang, Yiran and Zheng, Hao and Gong, Xuan and Li, Yihan and Liu, Lin and Liu, Zhuotao and Liang, Shiyu},
  title     = {{RLCracker}: Evaluating the Worst-Case Vulnerability of {LLM} Watermarks with Adaptive {RL} Attacks},
  booktitle = {Proceedings of the 43rd International Conference on Machine Learning},
  series    = {Proceedings of Machine Learning Research},
  volume    = {306},
  pages     = {48441--48483},
  year      = {2026},
  url       = {https://proceedings.mlr.press/v306/huang26ch.html}
}

@article{li2026trgof,
  author  = {Long, Qi and Su, Weijie J.},
  title   = {Robust Detection of Watermarks for Large Language Models Under Human Edits},
  journal = {Journal of the Royal Statistical Society Series B: Statistical Methodology},
  volume  = {88},
  number  = {2},
  pages   = {491--515},
  year    = {2026},
  doi     = {10.1093/jrsssb/qkaf056},
  url     = {https://doi.org/10.1093/jrsssb/qkaf056}
}

@inproceedings{diaa2025adaptive,
  author    = {Diaa, Abdulrahman and Aremu, Toluwani and Lukas, Nils},
  title     = {Optimizing Adaptive Attacks against Watermarks for Language Models},
  booktitle = {Proceedings of the 42nd International Conference on Machine Learning},
  series    = {Proceedings of Machine Learning Research},
  volume    = {267},
  pages     = {13546--13569},
  year      = {2025},
  url       = {https://proceedings.mlr.press/v267/diaa25a.html}
}

@inproceedings{bakshi2026openstamp,
  author    = {Bakshi, Miroojin and Rastogi, Saksham and Pruthi, Danish},
  title     = {{OpenStamp}: A Watermark for Open-Source Language Models},
  booktitle = {Conference On Language Models 3 pre-proceedings},
  year      = {2026},
  note      = {COLM 2026},
  url       = {https://proceedings.colm.cc/paper_files/paper/2026/hash/0e5cce15e1bfc6b3d7b71f24cc5da821-Abstract-Conference.html}
}

@article{christ2024provably,
  author  = {Christ, Miranda and Gunn, Sam and Malkin, Tal and Raykova, Mariana},
  title   = {Provably Robust Watermarks for Open-Source Language Models},
  journal = {arXiv preprint arXiv:2410.18861},
  year    = {2024},
  url     = {https://arxiv.org/abs/2410.18861}
}

@inproceedings{block2025gaussmark,
  author    = {Block, Adam and Rakhlin, Alexander and Sekhari, Ayush},
  title     = {{GaussMark}: A Practical Approach for Structural Watermarking of Language Models},
  booktitle = {Proceedings of the 42nd International Conference on Machine Learning},
  series    = {Proceedings of Machine Learning Research},
  volume    = {267},
  pages     = {4507--4569},
  year      = {2025},
  url       = {https://proceedings.mlr.press/v267/block25a.html}
}

@article{chen2024wapiti,
  author  = {Chen, Lingjie and Qiu, Ruizhong and Yuan, Siyu and Liu, Zhining and Wei, Tianxin and Yoo, Hyunsik and Zeng, Zhichen and Yang, Deqing and Tong, Hanghang},
  title   = {{WAPITI}: A Watermark for Finetuned Open-Source {LLM}s},
  journal = {arXiv preprint arXiv:2410.06467},
  year    = {2024},
  url     = {https://arxiv.org/abs/2410.06467}
}

@inproceedings{xu2025markyourllm,
  author    = {Xu, Yijie and Liu, Aiwei and Hu, Xuming and Wen, Lijie and Xiong, Hui},
  title     = {Mark Your {LLM}: Detecting the Misuse of Open-Source Large Language Models via Watermarking},
  booktitle = {1st Workshop on GenAI Watermarking, collocated with ICLR 2025},
  year      = {2025},
  url       = {https://openreview.net/pdf?id=JGTRj6h0Cv}
}

@inproceedings{zhao2025marktune,
  author    = {Zhao, Yizhou and Wu, Steven and Block, Adam},
  title     = {{MarkTune}: Advancing the Quality-Detectability Pareto Frontier of Open-Weight {LM} Watermarking},
  booktitle = {Lock-LLM Workshop at NeurIPS 2025},
  year      = {2025},
  url       = {https://openreview.net/forum?id=EYCC4yx3w4}
}

@inproceedings{xue2026pro,
  author    = {Xue, Jiaqi and Zhao, Yifei and Al Ghanim, Mansour and Gao, Shangqian and Sun, Ruimin and Lou, Qian and Zheng, Mengxin},
  title     = {{PRO}: Enabling Precise and Robust Text Watermark for Open-Source {LLM}s},
  booktitle = {NeurIPS 2026, Evaluations and Datasets Track},
  year      = {2026},
  url       = {https://arxiv.org/abs/2510.23891}
}

@article{feng2024certified,
  author  = {Feng, Xianheng and Liu, Jian and Ren, Kui and Chen, Chun},
  title   = {A Certified Robust Watermark for Large Language Models},
  journal = {arXiv preprint arXiv:2409.19708},
  year    = {2024},
  url     = {https://arxiv.org/abs/2409.19708}
}

@inproceedings{ren2023dimension,
  author    = {Ren, Jiaxiang and Zhou, Yang and Jin, Jiayin and Lyu, Lingjuan and Yan, Da},
  title     = {Dimension-Independent Certified Neural Network Watermarks via Mollifier Smoothing},
  booktitle = {Proceedings of the 40th International Conference on Machine Learning},
  series    = {Proceedings of Machine Learning Research},
  volume    = {202},
  pages     = {28976--29008},
  year      = {2023},
  url       = {https://proceedings.mlr.press/v202/ren23c.html}
}

@inproceedings{gloaguen2025durability,
  author    = {Gloaguen, Thibaud and Jovanovi{\'c}, Nikola and Staab, Robin and Vechev, Martin},
  title     = {Towards Watermarking of Open-Source {LLM}s},
  booktitle = {1st Workshop on GenAI Watermarking, collocated with ICLR 2025},
  year      = {2025},
  url       = {https://openreview.net/forum?id=OKjZvLJxOY}
}

@inproceedings{scharff2026merge,
  author    = {Scharff, Luisa and Gloaguen, Thibaud and Staab, Robin and Vechev, Martin},
  title     = {Making Open-Source Text {LLM} Watermarks Durable Against Merging},
  booktitle = {Advances in Neural Information Processing Systems (NeurIPS 2026)},
  year      = {2026},
  url       = {https://openreview.net/forum?id=84maDBcTYn}
}

@inproceedings{li2023quantization,
  author    = {Li, Linyang and Jiang, Botian and Wang, Pengyu and Ren, Ke and Yan, Hang and Qiu, Xipeng},
  title     = {Watermarking {LLM}s with Weight Quantization},
  booktitle = {Findings of the Association for Computational Linguistics: EMNLP 2023},
  pages     = {3368--3378},
  year      = {2023}
}

\end{document}